# Discovering Invariant Spatial Features in Electron Energy Loss Spectroscopy Images on the Mesoscopic and Atomic Levels

Kevin M. Roccapriore,[1] Maxim Ziatdinov,[1,2] Andrew R. Lupini,[1] Abhay P. Singh,[3,4] Usha Philipose,[3] and Sergei V. Kalinin[1]

[1]Center for Nanophase Materials Sciences, Oak Ridge National Laboratory, Oak Ridge, TN 37831
[2]Computer and Computational Sciences Division, Oak Ridge National Laboratory, Oak Ridge, TN 37831
[3]Department of Physics, University of North Texas, Denton, TX 76203
[4]Intel Corporation, Ronler Campus, Hillsboro, OR, 97124

Over the last two decades, Electron Energy Loss Spectroscopy (EELS) imaging with a scanning transmission electron microscope (STEM) has emerged as a technique of choice for visualizing complex chemical, electronic, plasmonic, and phononic phenomena in complex materials and structures. The availability of the EELS data necessitates the development of methods to analyze multidimensional datasets with complex spatial and energy structures. Traditionally, the analysis of these data sets has been based on analysis of individual spectra, one at a time, whereas the spatial structure and correlations between individual spatial pixels containing the relevant information of the physics of underpinning processes have generally been ignored and analyzed only via the visualization as 2D maps. Here we develop a machine learning-based approach and workflows for the analysis of spatial structures in 3D EELS data sets using a combination of dimensionality reduction and multichannel rotationally-invariant variational autoencoders. This approach is illustrated for the analysis of both the plasmonic phenomena in a system of nanowires and in the core excitations in functional oxides using low loss and core loss EELS, respectively. The code developed in this manuscript is open sourced and freely available and provided as a Jupyter notebook for the interested reader here.

# 1. Introduction

Electron energy loss spectroscopy (EELS) in a scanning transmission electron microscope (STEM) has become an enabling tool in studying electronic behaviors at the atomic and nanoscales. In a modern instrument, phenomena such as band gap, atomic elemental identification, and chemical nature of the specimen are all accessible.[1–4] Recently, monochromator technology has improved to the point where energy resolutions around 5 meV can be achieved, allowing vibrational spectroscopy to be conducted with unprecedented spatial resolution[5,6] in the near-infrared regime.[7,8] Effectively, this allows the electron microscope to explore atomic resolution phonon imaging, including single-atom and single-defect phonons, as well as identification of isotopes, plasmon-phonon coupling, and interfacial vibrational effects.[9–16]

Aside from vibrational spectroscopy, localized surface plasmon resonances (LSPRs) are also detectable and mappable in space.[17] Even more spatial detail can be realized by acquiring EELS images at a set of tilt angles, allowing a three-dimensional reconstruction of the plasmon modes which can serve to distinguish, for instance, modes localized to a surface from those at an edge.[18] Compared to traditional optical excitation schemes,[19,20] the electron beam simultaneously serves as an excellent excitation source and probe, in that it is a well-localized optically white light source whose energy transfer (e.g., to plasmon modes) can be spatially mapped with nanoscale precision. Given that LSPRs are typically supported in dimensionally confined systems like nanoparticles and nanowires, STEM-EELS emerges as the ideal tool to explore the local plasmonic behavior of these systems.

EELS may be performed as either a single point spectroscopy, or as a so-called "spectrum image," in which an EEL spectrum is collected at an array of equally spaced spatial coordinates. The latter is a three-dimensional dataset ('data cubes') containing information on energy dispersion and spatial distributions of relevant behaviors.[21] Correspondingly, the modern proliferation of EELS imaging necessitates development of methods to decipher physically meaningful information from this type of datasets. Note that this requirement is not unique to EELS and also emerges in the context of techniques such as energy-dispersive X-ray spectroscopy (EDS).[22,23]

Initial successes in exploring multidimensional EELS data sets were achieved with unsupervised linear unmixing methods, including principal component analysis (PCA), independent component analysis (ICA), and non-negative matrix factorization (NMF).[24] In these methods, the data set is represented as a linear combination of the components defined in the spectral domains and weights (or loading maps) defined in real 2D or 3D space as

$$R(x, y, E) = \sum_{i=1}^{N} A_i(x, y) r_i(E). \qquad \text{(Eq. 1)}$$

Here $A_i(x, y)$ represents the loading maps that contain spectral variability across the component space, and $r_i(E)$ are the spectral endmembers. The number of components, $N$, is chosen based on the anticipated physics and decomposition quality, among other factors.

Dissimilar linear decomposition methods impose different limitations on the weights and components. For example, PCA orders the components based on their significance and components are orthogonal; in NMF the components are non-negative, while techniques such as Bayesian linear unmixing impose an additional sum-to-one constraint.[25–27] Due to the constraint of non-negativity which is consistent with a counting spectroscopy, NMF has been used extensively in the electron microscopy community in order to spectrally deconvolve EELS data cubes,[28–30] while PCA and other algorithms without a non-negativity constraint tend to produce abstract, un-physical, and generally uninterpretable loadings.[31] Additionally, more significant constraints on sparsity, closeness to a chosen functional form, etc. can be imposed. In general, these constraints can be chosen to match the physics of the explored system with the hope that the resultant behaviors and loading maps will represent the physics of the system being explored.

Linear decomposition methods are particularly well suited for situations when the corresponding physical mechanisms are linear,[32,33] for example in a thin sample where the total core-loss EELS signal can be reasonably approximated as a linear combination of the signals from the atomic species in the probed volume. However, it is well recognized that in many cases this linear approximation does not hold, such as for multiple scattering in thick samples, or for features that change due to structural or bonding differences. Correspondingly, numerous methods have been developed for non-linear unmixing, or so-called "manifold learning." These methods ultimately seek to discover a low dimensional manifold containing the data based on similarity between the data points, statistical properties of the distribution, or partially known functional forms. Well-established techniques such as self-organized feature maps (SOFM),[34,35] Gaussian mixture modeling (GMM),[36] and simple and variational autoencoders[35,37,38] have existed for years in the computer vision fields but have only recently gained traction in the context of physics extraction. Beyond analysis of the data set to discover low dimensional representations, these approaches can be extended to establish structure-property relationships, e.g. predict spectral responses given a geometric input image and vice versa.[39] Generally, it is assumed that if the manifold containing the data is discovered, it provides a basis for the construction of generative models and can provide insight into the physics of the system.

However, the common limitation of both linear and manifold learning methods as applied to imaging data is that they are applied on the pixel-by-pixel level. In particular, we note that permutation of spatial pixels does not change the components and leads to an equivalent permutation of the loading maps. In this manner, the information contained in the spatial correlations within the image is generally lost. While a number of methods were suggested to impose spatial correlations in spectral analysis,[40–42] there has not been a universal approach to explore these in hyperspectral imaging methods. At the same time, spatial correlations in EELS and other hyperspectral techniques are a crucial aspect of the experiment, in which physics is naturally encoded, and is arguably a key motivating factor behind collecting a spectrum image.

To address this issue, we originally implemented invariant variational autoencoders[43] as a method for structural analysis of real-space images to disentangle rotation, translation, and scale from structural building blocks, allowing for definition of local symmetry.[44–48] Here, we introduce the *multichannel* rotationally invariant variational autoencoder (rVAE), which pairs classical dimensionality reduction with rotationally-invariant VAE such that the model inputs are still image-based, but are derived from a chosen spectral decomposition technique (e.g., NMF). The multichannel rVAE is demonstrated for EELS spectrum images and its applicability and performance in different EELS regimes is explored.

## 2. Experimental

As a first model system, we have chosen semiconductor nanowires. Nanowire systems support multiple plasmon modes and give rise to complex plasmonic interactions mediated by proximity effects. The energies of these depend both on geometry and effective dielectric properties (both the material and its environment), and in an extreme case, plasmons can even exist in molecular nanowires only a few atoms in length.[49] Moreover, the spatial localization of modes tends to decrease with lower energies, meaning there is a natural spatial structure that is related to the energy of the plasmon. As a second model system, we have chosen an example interface on a strontium titanate substrate for which we collected core-loss EELS spectra to explore the performance for atomically resolved data.

Indium antimonide nanowires of nominal diameter 100 nm were electrochemically synthesized[50] via a template assisted method (see Methods) and monochromated STEM-EELS was performed to acquire spectrum images of various configurations of nanowires. More details about sample growth, sample preparation, and STEM-EELS measurements can be found in the Methods section. A pair of nanowires with different lengths is selected, allowing for variability of the plasmon energies. Shown in **Figure 1** is the standard exploratory approach using NMF to spectrally deconvolve the data. At an array of probe positions enclosed within the red bounding box in Figure 1(a), EEL spectra were collected, and the resulting 3D data is flattened into two dimensions (one of space, one of energy). Figure 1 (b-d) illustrates that despite the lack of spatial details provided to the algorithm, NMF performs well in terms of extracting the different *spectral* features contained within the data, making it an excellent tool for spectral exploration within an EELS spectrum image, specifically in the case of nanoparticles and nanowires.[51] Several different plasmon modes are revealed by this deconvolution, where lower energy modes are more delocalized – the degree of localization to the nanowire is clearly depicted in the abundance maps in Figure 1(d).

In a different approach, we make use of a latent variable (nonlinear) technique – the VAE – to attempt to map the spectral features to a low dimensional manifold. Use of VAEs in hyperspectral imaging has been previously carried out in remote sensing fields[52,53] and recently have been used with STEM imaging,[45] where in both cases linear unmixing fails to capture the physics of the system. In contrast to linear unmixing methods, the autoencoder compresses the dataset to a small number of latent variables via a convolutional or fully connected neural network, and then expands it back to match the original signal with a different neural network in such a way as to retain the maximum original information. In this process, the relevant traits of the signal get encoded in continuous latent variables, whereas noise is rejected. In variational autoencoders, the additional constraint is the minimization of Kullback-Leibler divergence between encoded latent variables and a prior distribution.[54] Practically, this results in much smoother latent distributions in the latent space. However, both for AEs and VAEs, the general principle is that the data set is described in terms of continuous latent variables, with the encoder and decoder serving as functions establishing the relationship between the signal and latent vector. The distributions of the data in the low-dimensional latent space provide information on the variability of the behaviors within the system.

**Figure 2** shows what information is gleaned from application of the VAE to the identical data. A two-dimensional space was chosen for mapping the spectral features, where each encoded feature is represented as a latent variable. The latent variables elucidate the physical factors of variation in the data. A high dimensional latent space is usually not necessary as the prominent spectral features can be represented by a small number of latent variables. Practically, visualizing higher dimensional latent spaces is also difficult.

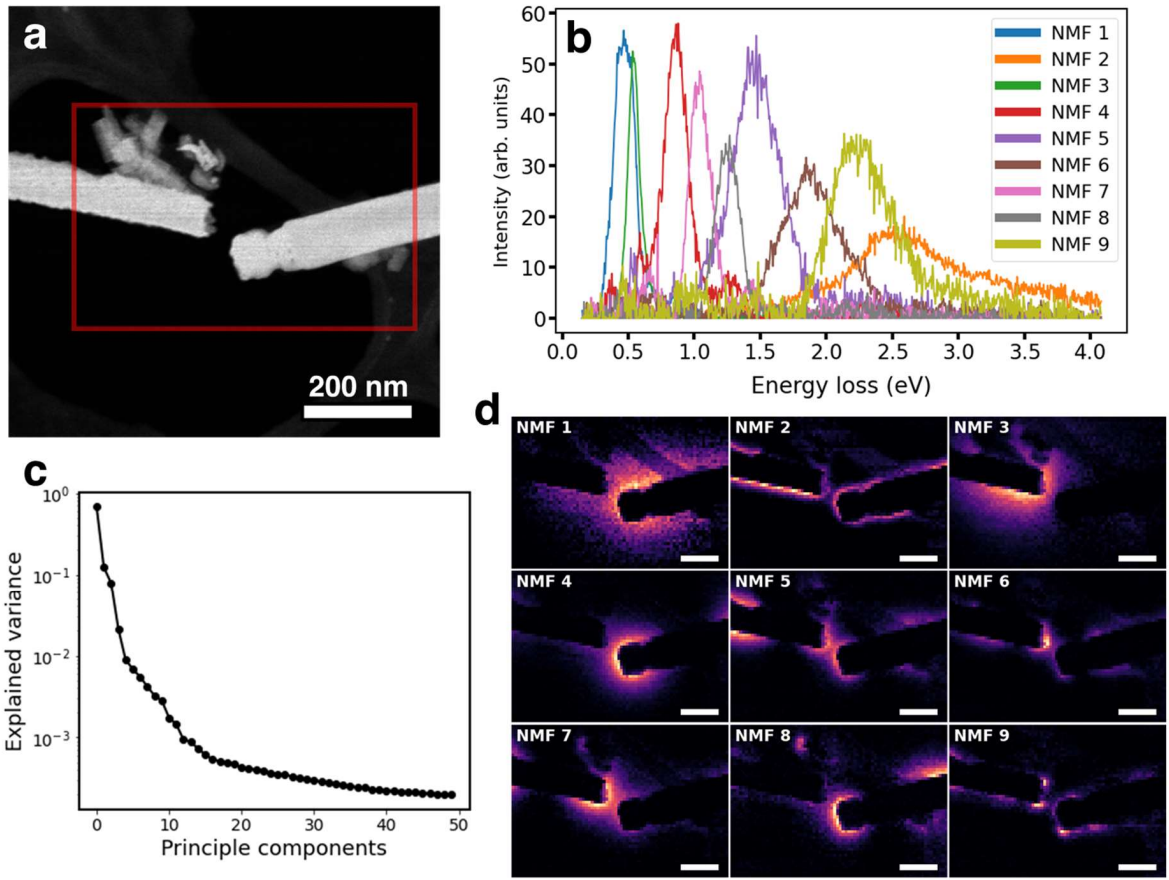

**Figure 1**. NMF deconvolution of an EELS spectrum image. A HAADF image is shown in (a) with the area used for the spectrum image outlined in red; (b) the NMF spectral endmembers; (c) PCA scree plot, representing the variability of the data; and (d) the corresponding abundance maps. Scale bars in (d) are all 100 nm.

The trained VAE model is used to encode the original data points into the latent space. Shown in Figure 2 (a,b) are the two latent feature maps, corresponding to the spatial distribution of the variability factors discovered by VAE. The distribution of points in latent space is shown in Figure 2 (c), overlaid with the kernel density estimation (KDE) which effectively allows visualization of a smoothed probability density using a Gaussian kernel.

The performance of the model is assessed in Figure 2 (d-f), where the model is used first to encode all spectra to a 2D latent space then to decode back to spectral space for each spatial coordinate, which effectively reconstructs the original data. We note that it is possible to have a favorable disentanglement (encoding) and a poor reconstruction (decoding).[55] Nevertheless, for the sake of simplicity we assess performance by considering these processes together. The energy-averaged original and decoded data are shown for comparison in Figure 2 (d) and (e), respectively, and (f) shows the mean squared error (MSE) for all energies (no averaging). The latter shows that the model has performed well, with mostly well below 1% error at any given position, which also reinforces our choice of two latent dimensions. Together, $Z_1$ and $Z_2$ contain the total - but compressed - spectral variability of the system.

By virtue of the compression strategy, noise is rejected during training, and as a direct result the data is therefore said to be denoised. The denoising can be seen when comparing the original and predicted energy-averaged spectrum images in Figures 2 (d) and (e), respectively. For a clearer insight into the spectral differences, we compare the spectra from a few pixels as in Figure 2 (g-i), where the model-reconstructed spectra clearly contain less noise than the original. In NMF and other linear methods, this noise is unavoidable.

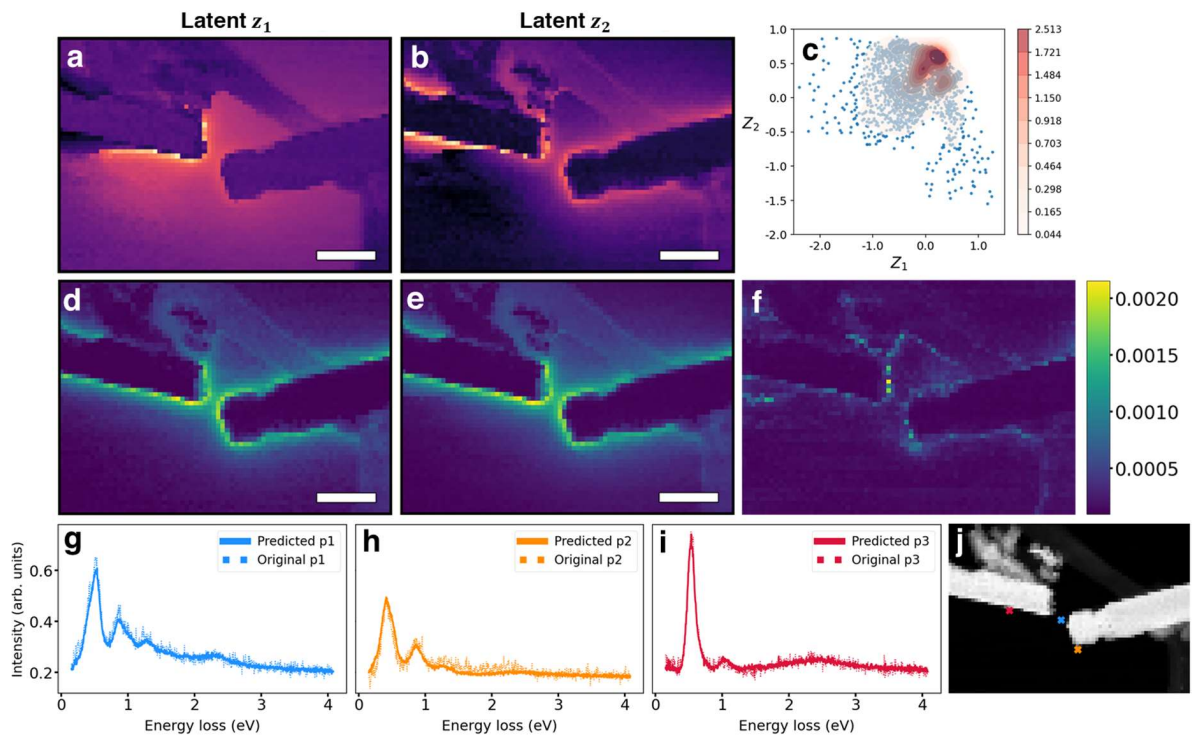

**Figure 2**. 1D VAE analysis. (a) and (b) represent the encoded latent feature images, while (c) shows the distribution of data points in the (2D) latent space overlaid with kernel density estimate (KDE). Spectral predictions for selected pixels shown in HAADF-STEM (j) are presented in (g-i). The hyperspectral data averaged over all energy is shown in (d), with the reconstructed image using the trained VAE model in (e), and the error surface in (f), with the MSE calculated for each pixel. Scale bars are 100 nm.

It is instructive to further compare the variational autoencoder with linear decomposition methods such as NMF and PCA. Like NMF, the VAE explores each spectrum without regard for its coordinate in space. Here our intent is primarily to discover spatial structures, but a challenge up until now has been treatment of a feature's rotation, as an object of interest may have any orientation in the image plane. To solve this problem, the rotational-invariant VAE (rVAE) encodes rotations as a separate latent variable.[45,56]

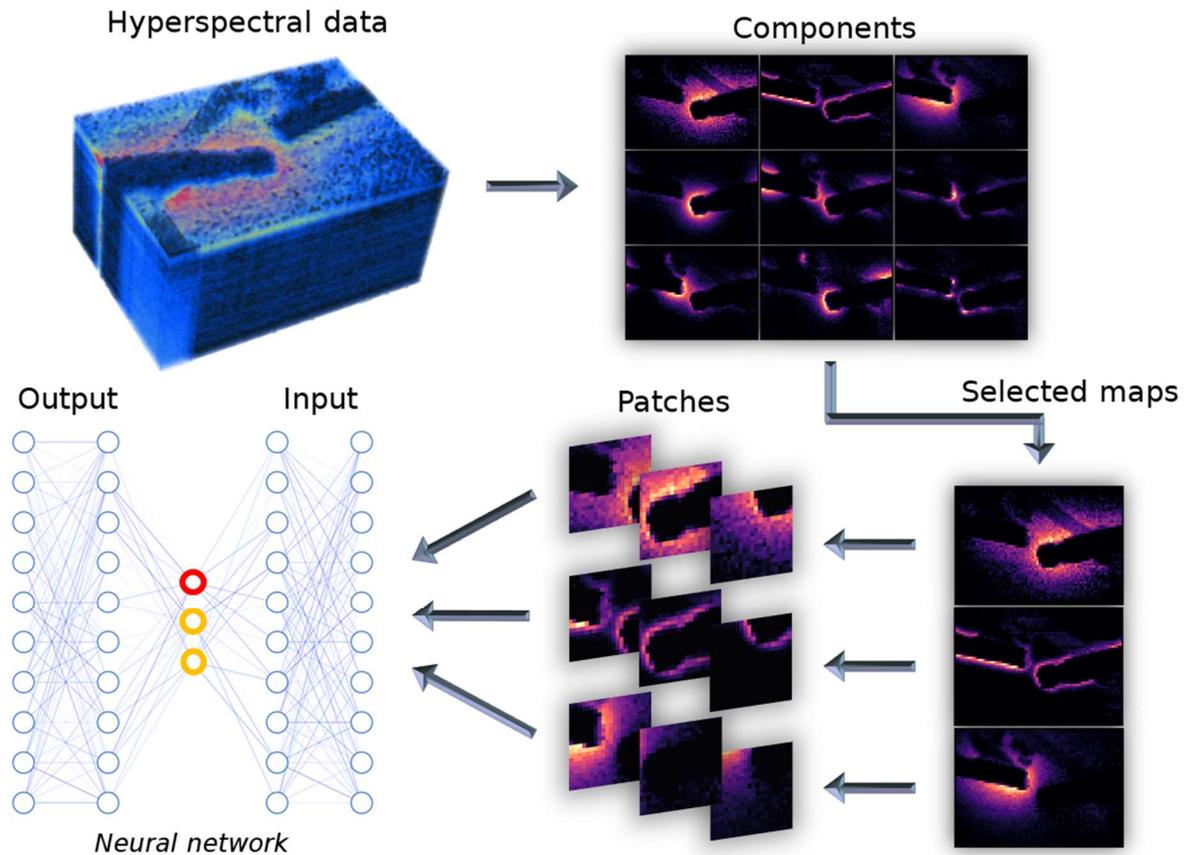

**Figure 3**. Multichannel rVAE design. A hyperspectral data cube is deconvolved to generate abundance maps based on spectral features. Using selected maps, sets of image patches are formed from each pixel in the parent map, then tiled into small windows (stacks of local channels). These multichannel tiles are sent as stacks to the rVAE network, which encodes all input to a latent space, from which latent features are subsequently decoded, iterating the process to best match output to input. The network shown here contains a latent space of three dimensions (red and yellow circles) where one dimension (red) is forced to represent the encoded angle.

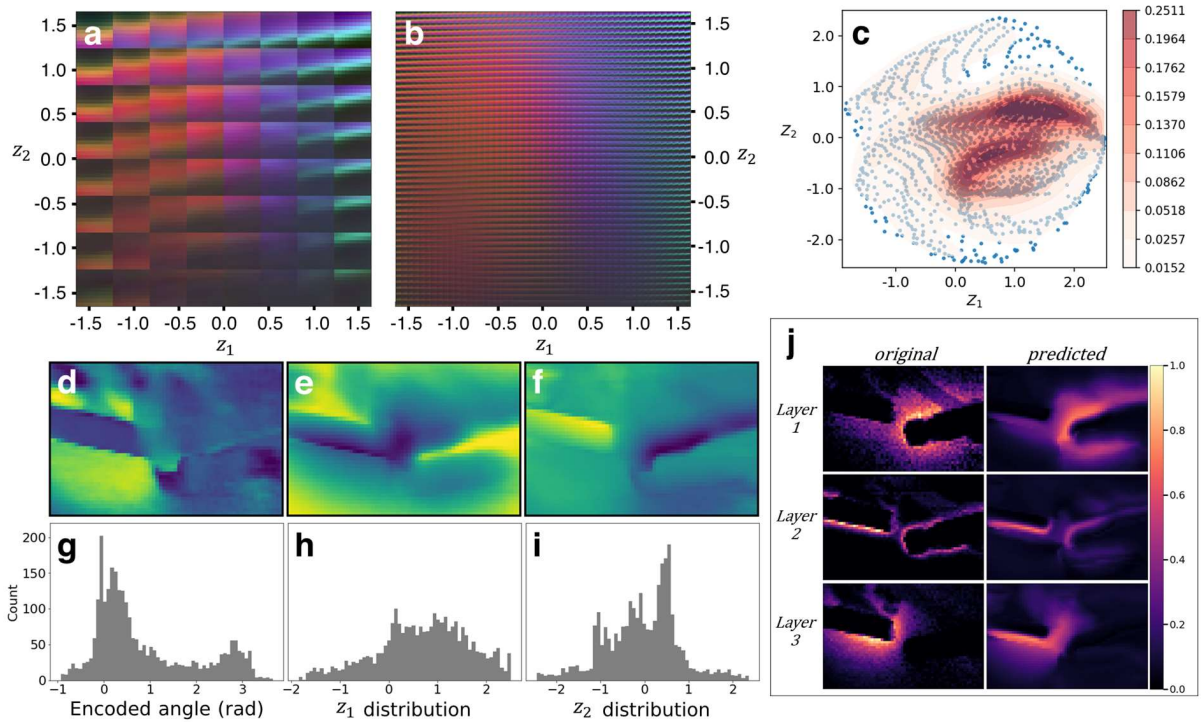

**Figure 4**. Multichannel rVAE performance for a window size of 10x10 pixels, 2 latent dimensions and 3 channels. (a) and (b) latent space representations are 2D manifolds, showing different voxel extremes. (c) Latent space distribution of the data overlaid with the kernel density estimate (KDE). (d) Angle, (e) latent feature map for $z_1$, (f) Latent feature map for $z_2$. Distribution of encoded angular features (g) and latent features $z_1$ (h) and $z_2$ (i) indicate the spatial variability. The selected three input channels are shown in (j) along with their predictions.

In the multi-channel rVAE architecture, a stack of images is used as inputs (and outputs). Images derived from NMF abundance maps are used here; therefore, each channel is associated with a particular NMF component and is based on a spectral feature intended to be explored. PCA maps or other spectral deconvolution methods can just as easily be used, however, we find that NMF tends to generally perform rather well for these EELS datasets and hosts the non-negativity constraint which physically agrees with a counting spectroscopy. We choose nine NMF components - out of a total of 30 – that best represent the system behavior as input channels, which are the same components from Figure 1. It is important to note that the choice of components falls in the realm of the domain expert, and while biased, it is necessary to produce a solvable problem.

With the aim of discovering spatial correlations that have a spectral foundation, a sub-image with size $dx \times dy$ is created at each pixel of each channel of a "reduced" spectroscopic image (pixels in which half the window size exceeds the boundary of the image are excluded). The window size plays a large role in the performance of the model and should be carefully chosen. We find that the model seems to perform well if the window size is commensurate with the characteristic length scale for the feature of interest – **Supplementary Figure S1** shows the effect of window size on the model performance. The stack of image patches for each channel are sent together to the rVAE algorithm and in a similar way to VAE, the algorithm attempts to minimize the difference between the decoded channels and original channels, (rather than a spectral evaluation) which at this stage is a purely image-based technique. Choice of number of channels and which channels to use is also left to the domain expert – in the case of this plasmonic nanowire system, one may choose to investigate the spatial structure of specific plasmon modes, e.g., a low-energy delocalized mode and a high-energy strongly localized mode, rather than all deconvolved components.

In **Figure 4**, we demonstrate the use of a 3-channel rVAE based on the first three NMF components, which incidentally includes both the lowest and highest energy plasmon modes that were observed. For consistency, we keep the number of latent dimensions the same as used in the 1D VAE case (two) and choose a window size of 10x10 pixels. The model used in Figure 4 reveals several features. First, there is variation in both latent space dimensions (e,f), while the encoded angle in (d) indicates edge behaviors – it delineates edges of the nanowires and variability of behaviors. Figure 4 (c) also indicates that the latent space does not collapse, indicated by a good distribution of points in latent space in both dimensions. This has the meaning that the data in the three chosen channels can be well-represented in a latent space of two dimensions. Conversely, if there is not a significant variation in one or more latent dimensions, it is said to have partially or fully collapsed. The distribution and occurrences of the latent features are also shown in histograms in (g-i). The character of the encoding can be represented by the latent space representation in (a,b). Here, the rectangular grid of points in the latent space is created and the corresponding image patches are decoded from these latent variable pairs. A 3-channel multichannel model is convenient in that each channel is assigned a color channel from the RGB color scheme – channel 1, red; channel 2, green; and channel 3, blue. Hence, the manifold maps in (a) and (b) represent the latent space distribution of effects relating to each of the three channels, where the colors signify the strength of that channel in the latent space. For example, the manifolds show a red color on the left-hand side and a blue color on the right most side, meaning the first channel's effects are strongest in the left region in latent space and the third channel is strongest in the right region. Green is hardly present which implies there is little strength and variation of the second channel. This component is extremely localized to the nanowire relative to the other two provided channels and so in most image patches the total intensity is considerably less, meaning that as far as the rVAE model is concerned, its effect is much less. Note that each channel is separately normalized, therefore the weights of the NMF components are intentionally lost. We typically found that if the set of channels are normalized together, there is substantially less variation in latent space and generally provides less insight, (likely) caused by an imbalance in the strength of the components. On the other hand, one may artificially increase the strength of this highly localized component (e.g., normalize from [0, 2]) to attempt to balance the component strengths, resulting in the green component – because it is the second channel – becoming visible in the manifold, as well as one of the latent features favoring edge behavior (**Supplementary Figure S2**). Having learned the variation, the model sufficiently predicts the three channels (Figure 4 (j)), even with only two latent features describing the system consisting of the chosen three channels.

One important model parameter that should be discussed in more detail is the size of window into which the input channels are divided. Like the number of latent variables and NMF components, this parameter should also be appropriately chosen to coincide with relevant spatial features. For strongly localized features, i.e., core loss EELS features which are localized to atoms or atomic columns, the window size should essentially be the size of an atom. However, for phenomena that tend to be more delocalized, the choice of window size needs to be explored more carefully. For instance, in the case of plasmons, a dipole mode may be much more delocalized than a higher order mode. If the data can be well-represented in two latent dimensions $z_1$ and $z_2$, there is no need to increase the complexity and extend to three or higher dimensions. However, this behavior is also data-dependent and should be explored.

As the number of channels to the multichannel rVAE is increased, the latent space distribution begins to take on a unique form. Reconstruction (prediction) of each channel also begins to improve, while the encoded rotation remains essentially unaffected. Analysis of latent space distribution with an increasing number of channels is presented in **Supplemental Figure S3**. The latent variables themselves change with almost every additional channel, which is not surprising given that each channel represents a different spectral feature (extracted via NMF), such that the information being compressed to the latent space is dynamically changing with each channel added to the model. Despite the increase of information, the latent space *distribution*, form, and encoded latent angles all remain constant, which is also striking.

It is important to emphasize the input to the multichannel rVAE is strictly image-based, but the images themselves have spectral origins from the spectral deconvolution method of choice. For this reason, choice of input channels is associated with the spatial features that will be learned. For example, choosing a 3-channel model with *different* NMF abundance maps for each model may produce a different result – this is studied further in **Supplementary Figure S4**. Consequently, when more channels are added, they represent more of the spectral (and spatial) variability contained in the data, and the output begins to converge and is less dependent on which channels are provided to the model. Rather than supplying the model with all components in an to attempt to account for all features of the system, it is generally preferrable to isolate specific features and unravel their physical impact and spatial correlations.

Switching energy regimes, we shift our focus to the performance of the VAE models on atomically resolved core loss EELS. Namely, we seek to understand ultimately if this method can be applied to spectral features that sustain a spatial variance on the *atomic* scale. First, **Figure 5** shows the 1D VAE - again with two latent variables - as applied to an atomically resolved spectrum image at a $SrTiO_3$-$LaGaO_3$ interface. The model extracts two distinct latent features $z_1$ and $z_2$ in Figure 5 (a) and (b) which clearly separate different interfaces, which is supported by the spread of points in latent space in (d). Comparison of the original and reconstructed average spectrum images in Figures 5 (e) and (f) provides almost an identical match, albeit with slightly less noise, resulting in a low MSE, as shown in (g). Two example pixels are selected to show individual spectral reconstructions and, as before, the channel-to-channel noise has been reduced. From these considerations and in terms of faithfully reconstructing the original data, we conclude the VAE model performed well. However, the problem remains that the 1D VAE ignores the spatial structure.

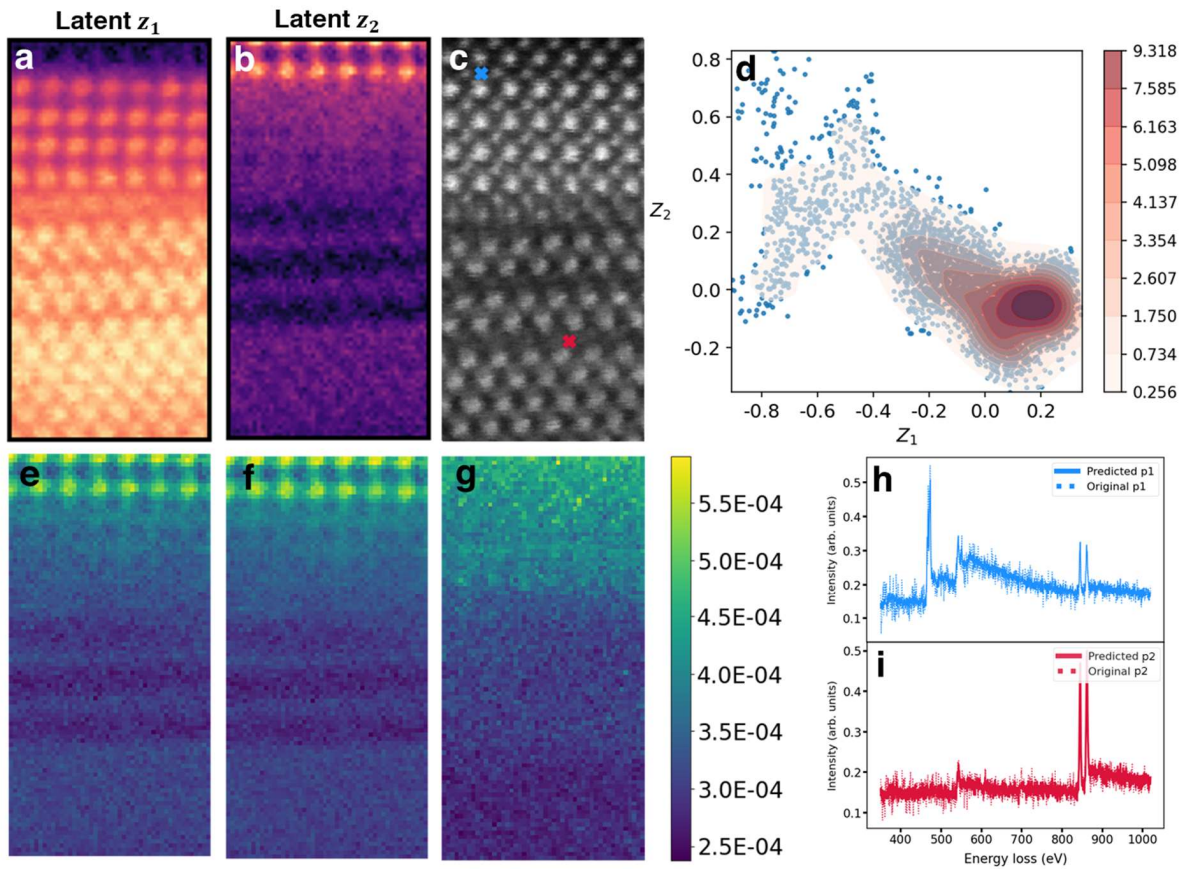

**Figure 5**. 1D VAE analysis for an atomically resolved core-loss EELS spectrum image of a $SrTiO_3$-$LaGaO_3$ interface. (a) and (b) are decoded latent features $z_1$ and $z_2$, respectively, with the HAADF-STEM image shown in (c). The distribution of points in latent space overlaid with the KDE is shown in (d). Ground truth spectrum image averaged over all energy and encoded image (reconstruction) after applying trained 1D VAE model shown in (e) and (f) respectively, where (g) is the MSE spatial map between (e) and (f). Selected spectra indicated in (c) and their VAE predictions are shown in (h) and (i).

To address this challenge, we apply a multichannel rVAE model with two latent variables and three input channels, as shown in **Figure 6**. The original and reconstructed channels are shown in (a-c), which, despite the presence of dense atomic features and a non-periodic structure, reveal that the atomic resolution contrast is still visible in the reconstructions. The encoded latent angle in (f) appears to again behave as an edge filter, but also reveals atomic contrast in the central part of the image that is not as clear in the input channels (or even the HAADF image). Comparing the latent variable maps from the rVAE (Figures 6 g,h) to those from the ordinary VAE (Figures 5 a,b) reveals their different performance. The VAE maps essentially show a bulk feature and a surface feature, while the rVAE distinguishes individual atomic channels. Finally, the color spread in RGB space in the manifold maps in Figure 6 (d,e) indicate each of the three inputs are mostly all utilized and relevant despite the second channel consisting mostly of low intensity. The distribution of points in latent space (Figure 6 (i)) takes on a peculiar structure where the KDE is clustered in two regions that represent the two preponderant compositions present at the interface.

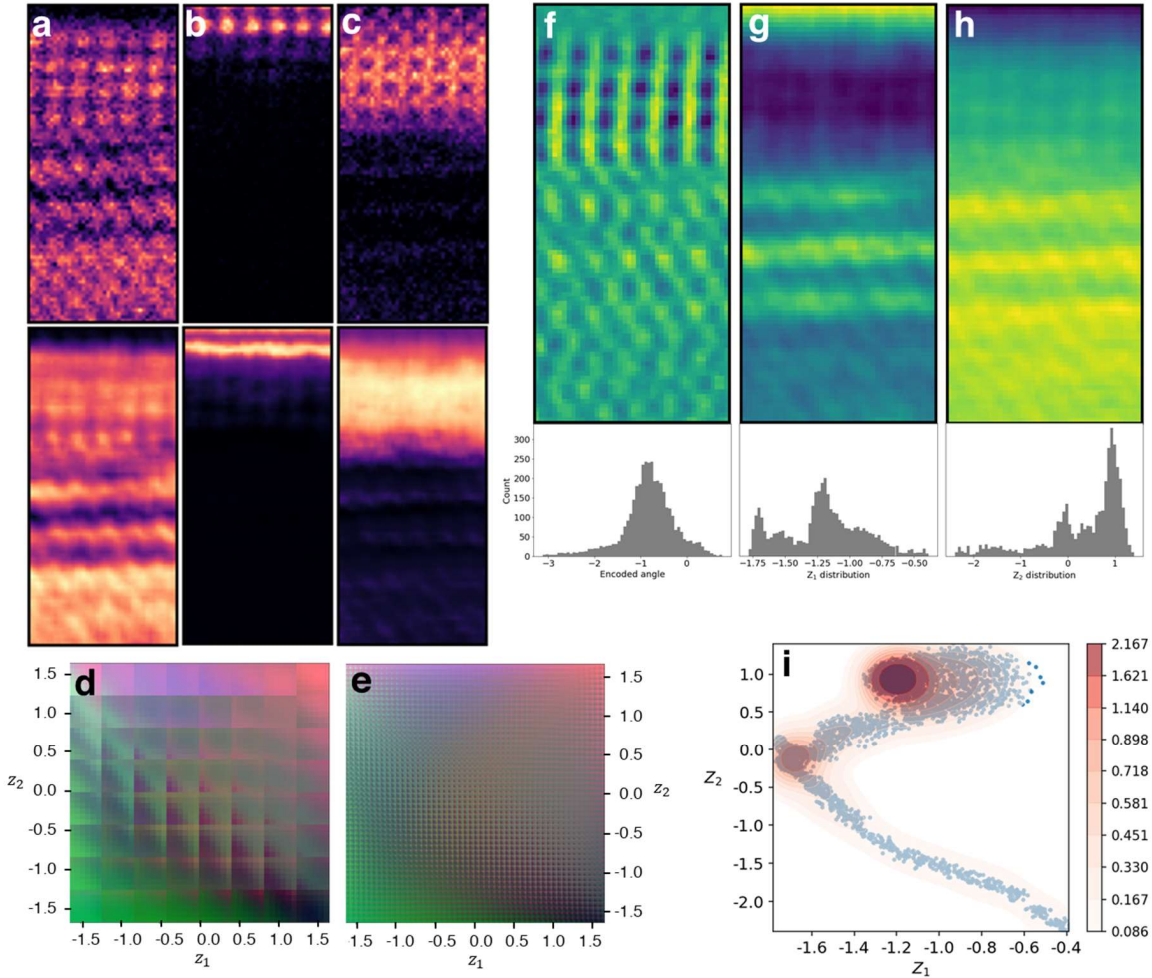

**Figure 6**. Multichannel (3 channel) rVAE analysis for atomically resolved core-loss EELS spectrum image, using shown NMF components as channels. Original channels are shown in (a-c) with their reconstructions in the panel directly below. Encoded latent rotation and latent features, $z_1$ and $z_2$ in (f-h) with distribution of occurrences. Manifold representations of latent space for different voxel extremes (d,e) and distribution of points in latent space (i) reveal distinct features.

To summarize, EELS and similar hyperspectral imaging methods yield information-rich data sets containing a wealth of information on materials structure and properties. Traditional approaches to extract features from hyperspectral images use linear or non-linear dimensionality reduction to explore the signal variability in the energy domain, while spatial details are extracted via visual examination of the loading or latent variable maps. Here we develop an approach to enable analysis of the spatial structures naturally embedded in hyperspectral images via rotationally invariant autoencoders. This approach allows identification of the elementary structures contained in the EELS data set while taking into account rotational invariances that are inevitably present in microscopic data.

This approach can be universally applied to hyperspectral imaging data in techniques such as continuous imaging tunneling spectroscopy (CITS) in scanning tunneling microscopy, electron-energy dispersive spectroscopy and cathodoluminescence in scanning electron microscopy, and multiple other hyperspectral and multichannel imaging techniques in electron, scanning probe, and optical imaging. We further note that development of these methods can proceed via configuring the latent space of the VAE, i.e., introducing the joint VAE concept[48] and control of the latent representations.[57]

**Acknowledgments**

This effort (ML and STEM) is based upon work supported by the U.S. Department of Energy (DOE), Office of Science, Basic Energy Sciences (BES), Materials Sciences and Engineering Division (K.M.R., S.V.K., A.R.L.) and was performed and partially supported (M.Z.) at the Oak Ridge National Laboratory's Center for Nanophase Materials Sciences (CNMS), a U.S. Department of Energy, Office of Science User Facility.

## Materials and methods

*Material synthesis*

InSb nanowires were grown commercially in a self-organized porus anodic aluminum oxide (AA0) template that was purchased from InRedox, that had a pore length and diameter of ~50 um and ~100 nm sequentially [1] and [2]. First, the template pores were expanded by it being soaked in 5% phosphoric acid ($H_3PO_4$) at 300° C for 5 minutes. Then, one side of the AAO template was coated with a thin gold (Au) film (~200 nm) using thermal evaporation (HHV Technologies, Linde: Auto 306). The template was exposed to pore wetting and de-aeration by having it placed in an ultrasonic bath in DI water for 2-3 minutes before the electrodeposition process.

The electrodeposition process was executed in a three-electrode cell in potentiostatic conditions at -1.5 V with the Au side of the AAO template functioning as a working electrode, platinum wire as an electrode and Ag/AgCl as the reference electrode. The electrolyte that was used was: 0.15 M $InCl_3$, 0.1 M $SbCl_3$, 0.36 M $C_6H_8O_7H_2O$, and $C_6H_5Na_3O_7$, and its pH value was adjusted to ~1.7. the AAO template was rinsed multiple times in DI water. dissolved from the template in 1 M KOH. In order for all traces of KOH to be removed, the nanowires were cleaned in DI water for the solution to be separated via centrifuge (Hermle model Z206A) at 3000 rpm for 2 minutes at 3000 rpm. Nanowires were then drop-cast onto a TEM grid for electron microscopy.

$SrTiO_3$-$LaGaO_3$ interface sample provided by Dr. C. Cantoni, with growth conditions described in [58]. Sample prepared for electron microscopy using standard cross-section and polishing techniques, followed by ion-beam milling to electron transparency.

*STEM – EELS*

Plasmon STEM-EELS experiments were collected using a 5$^{th}$ order aberration-corrected, monochromated STEM (Nion Hermes with Iris energy filter) operated at 60 kV with a 30 mrad semi-convergence angle and a nominal probe current of 25 pA. A dispersion of 10 meV was used. Spectrum images were acquired with pixel dwell times of 100 ms with an energy resolution of 40 meV defined by the FWHM of the zero-loss peak.

Core-loss STEM-EELS experiments were collected using an aberration-corrected Nion STEM operated at 100 kV with approximately 30 mrad convergence and collection angles. Spectrum images were acquired with pixel dwell times of 40 ms with an energy dispersion of 0.5 eV using a Gatan Enfina spectrometer.

## Data availability

The data used for analysis as well as additional materials are available through the Jupyter notebook located at: https://github.com/kevinroccapriore/Multichannel-rVAE/

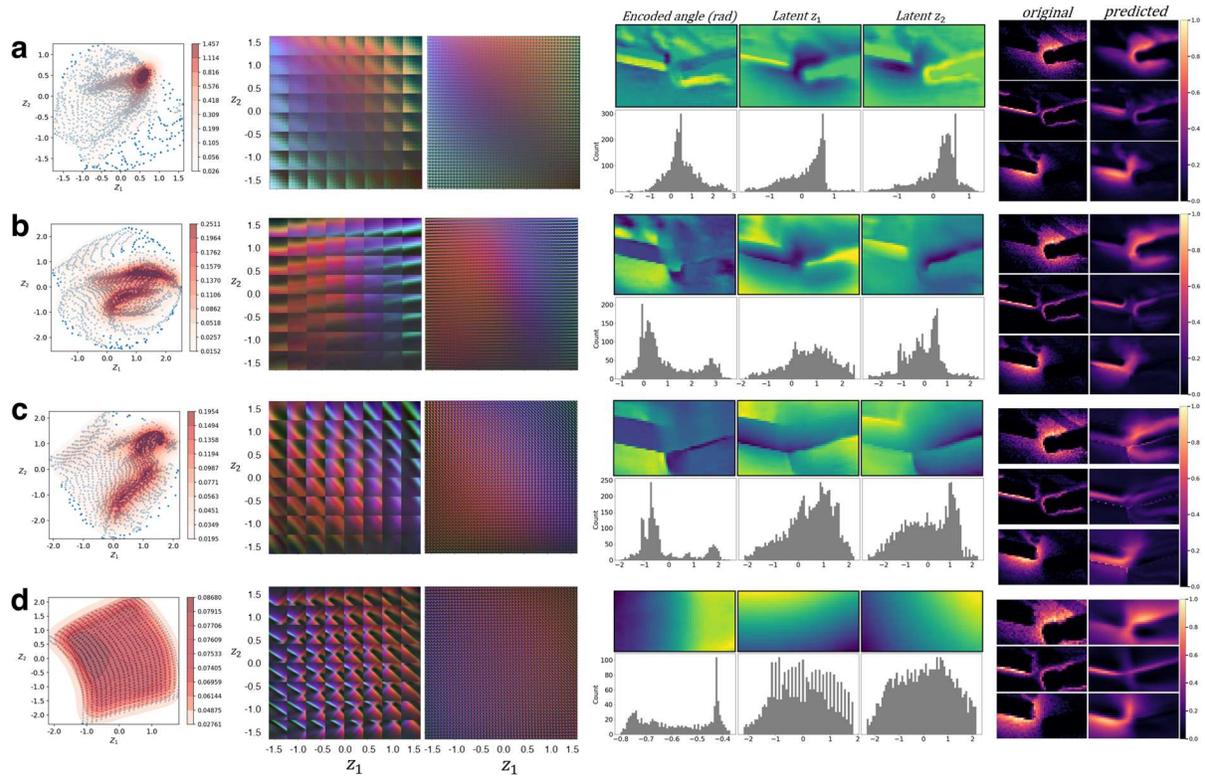

**Figure S1. Image patch relation to feature size**. Window sizes of 6, 10, 16, and 30 pixels are shown in (a-d) respectively, for the rVAE model architecture using the identical three input channels (first three NMF components) and training parameters. Panel (d) illustrates the latent features are dramatically smeared out (note the evenly distributed points in latent space, and the encoded latent feature maps), while at the other extreme of small window sizes, (a) shows much higher localization to the latent features (including rotational feature), as well as more localized clustering within the latent space.

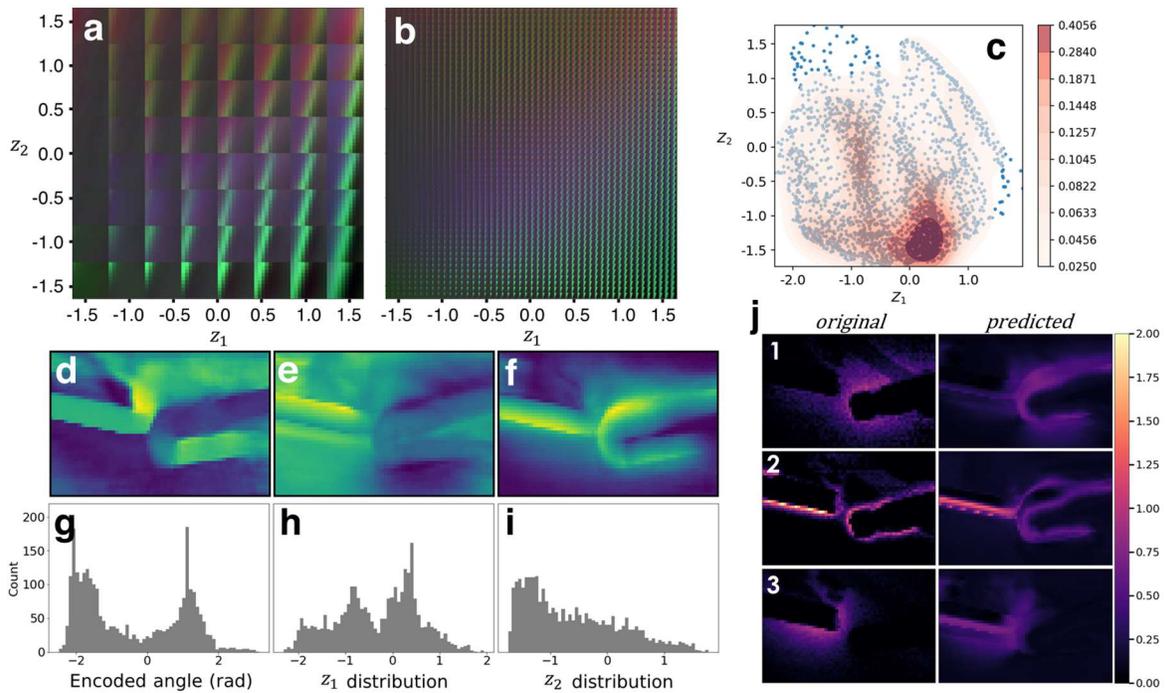

**Figure S2**. Weighting of specific channels. Here the model is trained identically to that of Figure 4, with the only difference being that the second channel is instead normalized to 2, enforcing a stronger effect from that channel. Being channel 2, the green channel now strongly dominates the latent space manifold representations in (a) and (b). Latent space distribution of data (c) overlaid with kernel density estimate (KDE). Three latent features are shown in (d-f) where (d) depicts the encoded angle in space, and (e,f) represent latent feature maps for latent dimensions $z_1$ and $z_2$, respectively. Distribution of angular features (g) and latent features $z_1$(h) and $z_2$ (i) indicate breadth of spatial variability. The selected three input channel are shown in (j) along with their predictions. Note the scale in (j), where channel 1 and 3 have maximum intensity of 1, while channel 2's maximum intensity is 2.

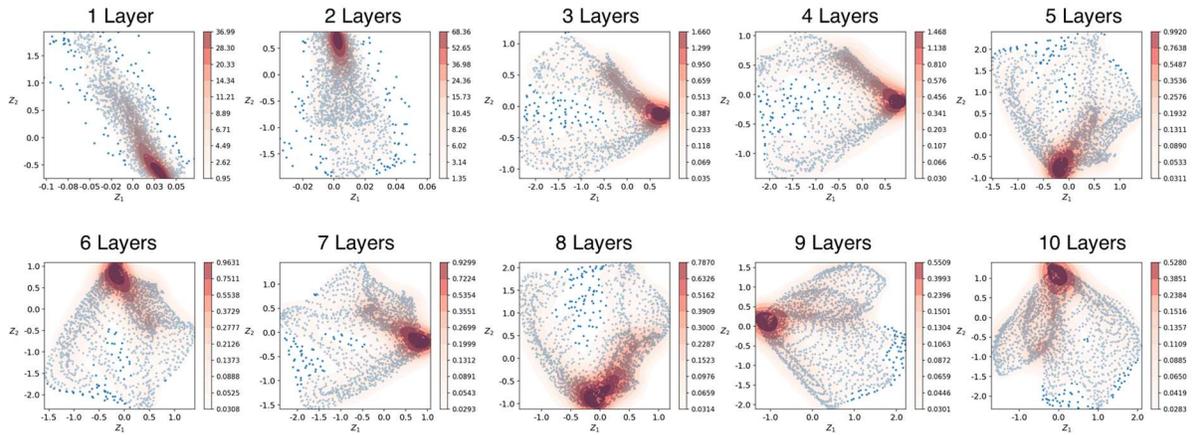

**Figure S3**. Latent space distribution for 2 latent variables as a function of number of channels. Note that for 1 and 2 channels, the latent coordinate $z_1$ has collapsed, indicating that 2 latent dimensions may not be needed to describe the system consisting of only 2 channels. For 3 or more channels, the points are distributed in a similar fashion – the apparent rotation of point distribution results from random initialization of model training and therefore does not represent additional behavior.

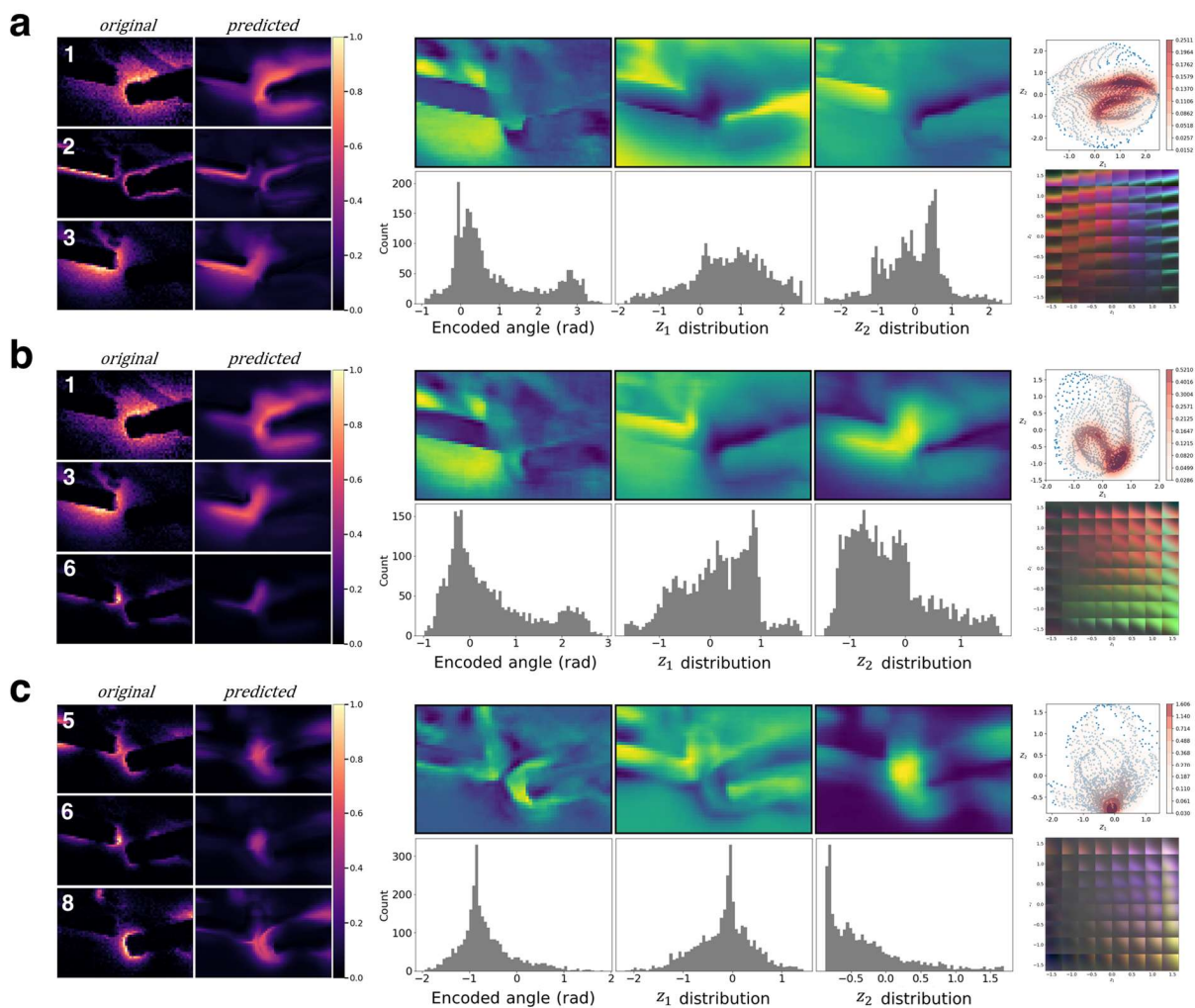

**Figure S4**. Choice of channel for 3-channel multichannel rVAE model. (a) shows the original first 3 NMF component channels, (b) shows using components 1,3, and 6, and (c) shows using components 5,6, and 8. The encoded latent variables, including the encoded angle, change with choice of channel, highlighting that the channels should be chosen based on features of interest.

23